\documentclass{article}
\usepackage{amsmath}
\DeclareMathOperator*{\AND}{\bigwedge}
\DeclareMathOperator*{\OR}{\bigvee}
\DeclareMathOperator*{\XOR}{\bigoplus}

\title{Boolean derivatives and computation of cellular automata}
\author{Franco Bagnoli\\
 Dipartimento di Matematica Applicata, \\
Universit\`a di Firenze,
Firenze, I 50139, Italy \\
also INFM and INFN, sezione di Firenze}
\begin{document}

\maketitle

\begin{abstract}
\noindent The derivatives of a Boolean function
are defined up to any order. The Taylor and MacLaurin expansions of a Boolean
function are thus obtained. The last corresponds to the ring sum
expansion (RSE) of a Boolean function,
and is a more compact form than the usual canonical
disjunctive form. For totalistic functions the RSE allows
the saving of a large number of Boolean operations.
The algorithm has natural applications to the
simulations of cellular automata
using the multi site coding technique.
Several already published algorithms are analized,
and expressions with fewer terms are generally found.
\\[.5cm]
\end{abstract}

\section{Introduction}
This work is based on the concept of
Boolean derivatives, introduced
in the context of cellular automata  by G. Vichniac~\cite{Vichniac}.
 A cellular automaton (CA)
is defined on a lattice by an interaction range
(for instance on a two dimensional square lattice, with nearest and
next to nearest neighbors interactions), and by an updating rule that
gives the future value (state) of a lattice variable given its present
state and the state of its neighbors. The rule is applied in parallel on
all the sites of the lattice, and can be either deterministic or
probabilistic. From a computational point of view, the simplest case
for the rule is a Boolean deterministic function, and, if not otherwise
specified, we shall refer to this situation in the following.
Cellular automata are often studied from a numerical point of view. Generally
large lattices and long time simulations are required, and this originates
the problem of developing
efficient algorithms for the simulations of cellular automata on
general purpose computers and sometimes on dedicated machines.
For the first
hardware resources, a technique that allows an efficient use of the
memory and CPU
is the Multi Site Coding (MSC) technique~\cite{fast,general,Stauffer}.
This technique
implies that the rule is coded only using bitwise operations.
Although  standard bitwise expressions (canonical forms)
are easy to generate
given a Boolean function,
the minimization of the number of required
operations is believed to be a np-complete problem~\cite{Wegener}.
In Section 2 we recall the basic definitions
and in the following section
we introduce the Boolean derivatives
that will lead to the Taylor and MacLaurin series for a  Boolean function.
They are
more compact expressions than the standard canonical conjunctive and
disjunctive forms.
In Section 4 this technique is farther developed
for the particular case of totalistic cellular automata,
for which the future state of a cell depends only on the total number of
neighbors that are in a certain state, regardless of their position.
The symmetries of the problem allow the saving of  a large number of Boolean
operations.
In Section 5 the results are applied to some models that appeared in
literature.
Finally, conclusions are drawn in the last section.

\section{General definitions}
Our fundamental set is
$B_1 = \{0,1\}$. This is called the Boolean set. Higher dimensional Boolean
sets are indicated as $B_n = \{0,1\}^n$. By ${\cal F}_{n,m}$ we
denote the set of functions $f : \{0,1\}^n \mapsto \{0,1\}^m$. ${\cal F}_n$
stands for ${\cal F}_{n,1}$.

To an element $a = (a_1, \dots, a_n) \in B_n$ corresponds a number
$a \in [0,2^n)$:
\[
    a = \sum_{i=1}^n a_i \cdot 2^{i-1};
\]
and to each number $a \in [0,2^n)$ corresponds a $n$-tuple $(a_1, \dots, a_n)
\in B_n$:
\[
	a_i = \left\lfloor a /2^{i-1} \right\rfloor \bmod 2,
\]
where $\lfloor a\rfloor$ stands for the integer part of $a$. An integer number is thus
an ordered set of Boolean numbers  (bits). In order to emphasize its Boolean
structure we shall refer to these sets with the name of {\it bitarray}, whose
dimension is that of the space in which it is defined.
We introduce a partial ordering between
bitarrays saying that $a \le b$ if, for all $i$, $a_i \le b_i$, where
$0 < 1$. We can thus substitute the expression $a \in [0,2^{n-1})$ by $a
\le 2^{n-1}-1$ or simply $a \in B_n$. A Boolean function $f$ is called
monotone if $a \le b$ implies $f(a) \le f(b)$.

This mapping between numbers and Boolean $n$-ple corresponds to
the representation of integer numbers in computers,
the integer division by a power of two being equivalent to
a right shift (in FORTRAN $\left\lfloor a /2^{i-1} \right\rfloor \equiv
{\tt SHIFTR(a,i-1)}$), and
the modulo two operation to take the leftmost (less
important) bit.

Let us introduce the most common Boolean operations. If applied to a bitarray,
they will act bit by bit. There are
$2^{2^n}$ Boolean functions in ${\cal F}_n$.
With $n=1$ the most important function is the negation (NOT),
indicated by the symbol $\lnot$, or by a bar over the argument. With
$n=2$ there are 16 functions. The ones that exist on (almost)
all computers are the AND, OR and
XOR operations. The AND (symbol $\wedge$)
gives one only if both the arguments are one
(it is equivalent to a multiplication
of the bits considered as integer numbers), the OR (symbol $\vee$) gives
one if either one or the other argument is one, while the XOR (symbol
$\oplus$) corresponds to the sum modulo two. Notice that the XOR operation
accounts both for the sum and for the subtraction or negation
($a \oplus a = 0$ and $a\oplus 1 = \lnot a$). The
AND has higher priority than the OR and XOR operations. The
negation has the highest priority. The OR and the XOR operations are
distributive with respect to the AND. The XOR operation can be expressed by
the NOT, AND and OR: $a \oplus b = \lnot a \wedge b \vee a \wedge \lnot
b$. Equivalently the OR can be expressed by the AND and XOR
operations: $a \vee b = a \oplus b\oplus a \wedge b$.
If two Boolean quantities $a$ and $b$ cannot be one at once, both
the expressions
$a \oplus b$ and $a \vee b$ give the same result. In the following we shall
emphasize this condition by writing $a + b$, and indeed in
this case the usual sum
operation can be used. On certain computers (namely on the
CRAY), the logical and numerical unities  can work in
parallel. By mixing Boolean and arithmetic operations it is
possible to speed up the actual calculations~\cite{Herrmann}.

Often in the literature~\cite{Wegener} the
conditional negation is indicated by $a^b$ with the meaning that $a^0 = \lnot
a$ and $a^1 = a$. This is equivalent to $\lnot (a \oplus b)$ or to
$a \oplus b \oplus 1$. In this work we prefer to assign a different meaning to
exponentiation, more similar to the usual one.
We define $a^0 = 1$ and $a^1 = a$. With this definition the expression
$a^b$ is equivalent to $(a\oplus 1)\wedge b \oplus 1$ or to $a \vee \lnot b$.
 When applied to bitarrays, the exponentiation is performed bit by bit
and the results are afterwards ANDed together,
\begin{equation}
a^b = \AND_{i=1}^n a_i^{b_i} = \AND_{i=1}^n a_i\vee \lnot b_i. \label{exp}
\end{equation}
We note that $a^b$ is equal to one if and only if $a \le b$.

A Boolean function $f\in {\cal F}_n : x\in B_n \rightarrow f(x)$
is defined by giving its results for all the possible ($2^n$) configurations
of its arguments (truth table). It is possible to obtain an
expression for the $f$ only containing the AND, OR and NOT operations
from this table.
It is sufficient to give $f^{-1}(0)$ or
$f^{-1}(1)$. The two canonical forms are
\[
\begin{split}
	f(x) &= \OR_{ a \in f^{-1}(1)} \quad\AND_{i=1}^n \lnot(x_i\oplus a_i);
    \qquad \hbox{disjunctive form, and}\\
    f( x) &= \AND_{ a \in f^{-1}(0)} \quad\OR_{i=1}^n {x_i \oplus a_i};
    \qquad \hbox{conjunctive form.}
\end{split}
\]

These canonical forms are the standard starting points for the problem of
minimizing	the number of required operations given a function.
It is possible to demonstrate~\cite{Wegener} that the
minimal expression for a monotone function only contains the AND and OR
operations; the expression is unique and easy to compute.

As the NOT and the OR can be expressed by  the AND and
XOR operations, any function $f$ can be given in terms of
the latter two operations (Ring Sum Expansion, RSE)~\cite{Wegener}.
This form is identified by a
Boolean vector $f_i; i \in B_n$
\begin{equation}
    f(x) = \XOR_{i \in B_n} \quad f_i
	\wedge x^i.	 \label{rse}
\end{equation}

Since the number of different Boolean vectors  $f_i$
and of functions $f\in {\cal F}_n$ is equal, the RSE is unique.

\section{Boolean derivatives}
Following Vichniac~\cite{Vichniac}, we define the derivative of a
Boolean function
$f\in {\cal F}_n$ with respect to its $i$-th argument $x_i$
as
\[
   \frac{\partial f}{\partial x_i}\bigg|_x = f(x_1,\dots,x_i,\dots,x_n) \oplus
	  f(x_1,\dots,\lnot x_i,\dots,x_n).		\label{deriv}
\]

This (first order) derivative expresses the dependence of the function by its
$i$-th argument: $\partial f/\partial x_i$ is one if $f$ changes when
changing
$x_i$, given the configuration $x_1,\dots,x_{i-1},x_{i+1},\dots,x_n$. If the
derivative of $f$ with respect to $x_i$ is one regardless of the other
arguments, than the rule changes its value whenever $x_i$
does. In Ref.~\cite{Ien} a rule that shows this behavior is called a toggle rule.

This definition is consistent with the common expectations:
 the derivative of the identity function is
one, and the derivative of a constant (0 or 1) is zero. Moreover, the
derivative is linear with respect to the XOR operations, and it follows the
standard rule for the derivative of a product,
\[
    \frac{\partial (f\wedge g)}{\partial x} =
		 \frac{\partial f}{\partial x} \wedge g \oplus f \wedge
		  \frac{\partial g}{\partial x}.
\]

We can extend the definition to higher order derivatives. For example, a
second order derivative with respect to $x_i$ and $x_j$ is defined as
\[
\begin{split}
	\frac{\partial ^2 f }{\partial x_i\partial x_j}\bigg|_x = {}&f(x_1,\dots,x_i,\dots,x_j,\dots,x_n)
     \oplus f(x_1,\dots,\lnot x_i,\dots,x_j,\dots,x_n) \oplus \\
		   &\quad f(x_1,\dots,x_i,\dots,\lnot x_j,\dots,x_n) \oplus 
		f(x_1,\dots,\lnot x_i,\dots,\lnot x_j,\dots,x_n).
\end{split}
\]

Note that the definition is consistent with the usual chain rule for
derivatives, i.e.,
\begin{equation}
     \frac{\partial ^2 f}{ \partial x\partial y}=
		 \frac{\partial }{\partial y}\left(\frac{\partial f}{\partial x}\right).
		 \label{chain}
\end{equation}

A second order derivative with respect to the same argument is
identically zero.

We introduce a more compact definition for the Boolean
derivatives. Indicating with $x$ the bitarray of the arguments
$(x_1,\dots,x_n)$ and with $\delta$ a (constant) bitarray of the same
dimension, we define
\[
    \partial _\delta f(x) = \XOR_{\alpha\le \delta} f(x \oplus \alpha).
\]
It is immediate to verify that $\partial _\delta f(x)$ is equal to the partial
derivative in $x$ of $f$ with respect to the variables
that correspond to nonzero bits in $\delta$.
For instance, indicating with $\delta^{(i)}$ a bitarray in which only the
$i$-th bit is set to one (i.e., $\delta^{(i)}_i = \delta_{i,j}$, where the
latter is the usual Kronecker symbol), we have
\[
    \partial _{\delta^{(i)}} f(x) = \frac{\partial f }{ \partial x_i}\bigg|_x.
\]

We can now state our most important result. For a Boolean function
the Taylor expansion is always
finite. Let us start with a perturbation on only one variable.
 If $y = x \oplus \delta^{(i)}$, from the definition~\eqref{deriv} of the
derivative we get
\[
    f(y) = f(x) \oplus \partial _{\delta^{(i)}} f(x).
\]

Generalizing
\[
	f(x\oplus\delta) = \XOR_{\alpha \le \delta} \partial _\alpha f(x),
\]
with $\partial _0f(x) = f(x)$.

Using our definition of the exponentiation~\eqref{exp}, we can
substitute the XOR over $\alpha \le \delta$ with a
XOR over $\alpha \in B_n$. We need a test function that gives one if
$\alpha \le \delta$ and zero otherwise,
and from the consideration after Eq.~\eqref{exp} this
can be expressed as $a^b$.
Finally we obtain
\[
    f(x\oplus\delta) = \XOR_{\alpha \in B_n}
     \delta^\alpha \partial _\alpha f(x).
\]

As a noticeable consequence, we can expand a function starting from 0 (the
bitarray that has all the bits to zero), obtaining the MacLaurin series
\begin{equation}
    f(x) =  \XOR_{\alpha \in B_n} x^\alpha \wedge f_\alpha,
\label{expansion}
\end{equation}
where
\begin{equation}
f_\alpha = \partial _\alpha f(0);\label{falpha}
\end{equation}
Which is the ring sum expansion
of the function $f$, Eq.~\eqref{rse}.

Let us explicitly write down the formula~\eqref{expansion} for
 an elementary CA, whose evolution rule depends on the cell itself ($y$) and on its
nearest neighbors ($x$ and $w$). Locally
\[
    y' = f(x,y,w)
\]
 where the prime indicates the future value of the cell. The MacLaurin
 expansion of $f$ is given by
\[
\begin{split}
	y' = {}&f(0,0,0) \oplus \\
	   &\quad x \wedge \frac{\partial f}{ \partial x}\bigg|_{0,0,0} \oplus y \wedge 
		 \frac{\partial f}{
	\partial  y}\bigg|_{0,0,0}	\oplus z \wedge \frac{\partial f}{ \partial z}\bigg|_{0,0,0} \oplus\\
	&\quad \quad x\wedge y\wedge \frac{\partial f^2}{ \partial x\partial y}\bigg|_{0,0,0} \oplus
	x \wedge z \wedge \frac{\partial f^2}{\partial x\partial z}\bigg|_{0,0,0} \oplus
	y\wedge z \wedge \frac{\partial f^2}{ \partial y\partial z}\bigg|_{0,0,0} \oplus \\
	&\quad \quad \quad x\wedge y\wedge z\wedge \frac{\partial f^3}{ \partial x\partial y\partial z}\bigg|_{0,0,0}.
\end{split}
\]

The first order derivatives of all the elementary CA
can be found in Ref.~\cite{Vichniac}.
Higher order derivatives can be obtained by using the chain rule~\eqref{chain}.
Otherwise, the array of derivatives $f_i$ in zero
can be obtained from the truth table $f(j)$ via the matrix
${\cal M}_{i,j}$
\[
 f_i = \XOR_{j \in B_n} {\cal M}_{i,j} \wedge f(j);	
\]
where
\[
{\cal M}_{i,j} = \binom{j}{i} \mod 2.
\]

The matrix $\cal M$ can be recursively generated considering that
\begin{equation}
\begin{split}
{\cal M}_{i,0} &= 1;\\
{\cal M}_{0,j} &= 0 \quad (j> 0);\\
{\cal M}_{i,j} &= {\cal M}_{i-1,j} \oplus {\cal M}_{i-1,j-1} \quad (i,j > 0);
\end{split}
\label{defM}
\end{equation}

To show an application of the MacLaurin expansion, let us examine the
expression normally used to select between two random bits
$a$ and $b$ according to a third one ($r$),
\[
	f(r) = r \wedge a \vee \lnot r \wedge b \qquad \hbox{(4 operations)}.
\]
We only consider the explicit dependence of the function on $r$.
To write down the ring sum expansion of $f(r)$ we need
\[
\begin{split}
    f(0) &= b;\\
    \partial _1 f(0) &= f(0) \oplus f(1) = b\oplus a.
\end{split}
\]

The RSE for $f(r)$ is
\[
    f(r) = b \oplus r \wedge (a \oplus b) \qquad \hbox{(3 operations)}.
\]
We consider it a good result
to save one operation out of four in such a widely
used and (apparently) simple expression. Other examples can be found in
section~5.
\catcode`"=\active
\def"{^{(n)}}

\section{Totalistic rules}
The power of the algorithm is particularly evident when applied to totalistic
CAs. The transition rule for these
automata depends on the sum of the cell values in the
neighborhood,
\[
T" = \sum_{i=1}^n x_i.  
\]

Any totalistic evolution rule can be written as 
\begin{equation}
f(x_1,\dots,x_n) = f\left(T"\right) = \sum_{k=0}^9 {r_k\wedge\chi"_k}
\label{totalistic}
\end{equation}
where $\chi"_k$ is one if $T" = k$
and zero otherwise (totalistic characteristic functions).
Only one term contributes in the sums of
equations~\eqref{totalistic} so that we can use the arithmetic summation.
 The quantities $r_k$
take the value zero or one and define the automaton rule.
Probabilistic CAs may be implemented by allowing the
coefficients $r_k$ of equation~\eqref{totalistic} to assume
the values zero and one with  probabilities $p_k$ (see also
the last example of Section~5).

A totalistic function $f$ is completely symmetrical
with respect to its arguments~\cite{symmetric}. This implies that
the derivatives of $f$ of same order are all functionally equals. In
particular, as the derivatives of the MacLaurin expansion~\eqref{expansion} are
calculated in zero, they are actually equals, and thus can be factorized.
This leads to
\begin{equation}
  f(x_1,\dots, x_n) = f\left(T"\right) =
  f_{0} \oplus f_{1} \wedge \xi"_1 \oplus f_{2} \wedge \xi"_2
  \oplus \dots f_{n} \wedge \xi"_n; \label{totalexpansion}
\end{equation}
where the $f_i$ represents the	derivative of order  $i$ of $f$ in
0~\eqref{falpha}, and the $\xi"_i$
are the homogeneous polynomes of degree $i$ in the variables $x_1,\dots, x_n$
(using the AND for the multiplication and the XOR for the sum)
\[
\begin{split}
	\xi"_1 &= x_1 \oplus x_2\oplus \dots \oplus x_n,\\
	\xi"_2 &= x_1 \wedge x_2 \oplus x_1 \wedge x_3 \oplus \dots \oplus x_{n-1} \wedge
      x_n,\\
	&\dots\\
	\xi"_n &= x_1 \wedge x_2 \wedge \dots\wedge x_n.
\end{split}
\]

The functions $\xi_i$ satisfy some recurrence relations. The first one is
based on the idempotent property of the AND operation ($a\wedge a = a$)
and the nullpotent property of the XOR operation ($a \oplus a = 0$)
\[
\begin{split}
    \xi_1 &: \hbox{ irreducible;}\\
    \xi_2 &: \hbox{ irreducible;}\\
    \xi_3 &= \xi_2 \wedge \xi_1;\\
    \xi_4 &: \hbox{ irreducible;}\\
    \xi_5 &= \xi_4 \wedge \xi_1;\\
    \xi_6 &= \xi_4 \wedge \xi_2;\\
    \xi_7 &= \xi_4 \wedge \xi_3 = \xi_4 \wedge \xi_2\wedge \xi_1;\\
    \xi_8 &: \hbox{ irreducible;}\\
	&\dots\ .
\end{split}
\]
The second property is based on the separation of the variables in two
groups (bisection). Let us call $X$ the group of the variables $(x_1,\dots, x_n)$,
with $L$ we indicate the left part of $X$ up to some index $j$, and
with $R$ the right part of $X$
\[
\begin{split}
	L &= (x_1,\dots, x_j)\\
	R &= (x_{j+1},\dots, x_n).
\end{split}
\]

We have
\[
\begin{split}
\xi_i(X) = {}&\xi_i(L) \oplus \xi_{i-1}(L)\wedge \xi_1(R) \oplus
	\xi_{i-2}(L)\wedge \xi_2(R) \oplus \dots \\
	   &\quad\oplus\xi_1(L)\wedge \xi_{i-1}(R) \oplus \xi_i(R).
\end{split}
\]

As an example, let us explicitly calculate the $\xi_i$ for eight
variables. We bisect homogeneously the set $X = (x_1,\dots,x_8)$
first into $L$, $R$, and then
into $LL$, $LR$, $RL$, $RR$. We have
\[
\begin{cases}
\xi_1(LL)  &= x_1 \oplus x_2,\\
\xi_1(LR)  &= x_3 \oplus x_4,\\
\xi_1(RL)  &= x_5 \oplus x_6,\\
\xi_1(RR)  &= x_7 \oplus x_8;\\
\xi_2(LL)  &= x_1 \wedge x_2, \\
\xi_2(LR)  &= x_3 \wedge x_4, \\
\xi_2(RL)  &= x_5 \wedge x_6, \\
\xi_2(RR)  &= x_7 \wedge x_8;\\
\end{cases}
\]
\[
\begin{cases}
\xi_1(L) &= \xi_1(LL) \oplus \xi_1(LR), \\
\xi_1(R) &= \xi_1(RL) \oplus \xi_1(RR);\\
\xi_2(L) &= \xi_2(LL) \oplus \xi_1(LL) \wedge \xi_1(LR) \oplus
\xi_2(LR),\\
\xi_2(R) &= \xi_2(RL) \oplus \xi_1(RL) \wedge \xi_1(RR) \oplus
\xi_2(RR);\\
\xi_3(L) &= \xi_2(L) \wedge \xi_1(L), \\
\xi_3(R) &= \xi_2(R) \wedge \xi_1(R);\\
\xi_4(L) &= \xi_2(LL)\wedge \xi_2(LR), \\
\xi_4(R) &= \xi_2(RL)\wedge \xi_2(RR);\\
\end{cases} 
\]
\[
\begin{cases}
\xi^{(8)}_1 &= \xi_1(L) \oplus \xi_1(R); \\
\xi^{(8)}_2 &= \xi_2(L) \oplus \xi_1(L)\wedge \xi_1(R) \oplus \xi_2(R);\\
\xi^{(8)}_3 &= \xi_2 \wedge \xi_1;\\
\xi^{(8)}_4 &= \xi_4(L) \oplus \xi_3(L) \wedge \xi_1(R) \oplus \xi_2(L)\wedge
  \xi_2(R)
  \oplus \xi_1(L) \wedge \xi_3(R) \oplus \xi_4(R);\\
\xi^{(8)}_5 &= \xi_4 \wedge \xi_1;\\
\xi^{(8)}_6 &= \xi_4 \wedge \xi_2;\\
\xi^{(8)}_7 &= \xi_4 \wedge \xi_3;\\
\xi^{(8)}_8 &= \xi_4(L) \wedge \xi_4(R).
\end{cases}
\]
where $\xi^{(8)}_k = \xi_k(X)$.
Taking into account the common patterns that appear in the expressions of
$\xi^{(8)}_2$ and $\xi^{(8)}_4$, we only need 34 operations to build
up all the $\xi^{(8)}_i$.

The extension of the calculations to 9 variables,
only adds a small number (16) of operations
\[
\begin{split}
\xi^{(9)}_1 &= \xi^{(8)}_1 \oplus x_9, \\
 \xi^{(9)}_i &= \xi^{(8)}_i \oplus
\xi^{(8)}_{i-1}    \wedge x_9
\quad (2 \le i\le 8), \\
\xi^{(9)}_9    &= \xi^{(8)}_8 \wedge x_9;
\end{split}
\]
even though a careful bisection of the set of variables implies fewer (39)
operations.

A kind of normalization condition on the $\xi_i$ is given by
\begin{equation}
\OR_{i=1}^n x_i = \XOR_{i=1}^n \xi"_i, \label{or}
\end{equation}
and can be used to save operations in building an expression containing a
XOR of $\xi_i$. Another useful relation is
\begin{equation}
	\OR_{i=1}^{n-1} (x_i \oplus x_{i+1}) = \XOR_{i=1}^{n-1} \xi"_i. 
	\label{or2}
\end{equation}

We now have to build up the derivatives (in zero) of a totalistic function
$f$, Eq.~\eqref{totalexpansion}. There are only $n+1$ independent derivatives
$f_i$, $(i=0,\dots,n)$,  as
all the derivatives of the same order $i$ are equal.
We have
\[
f_i = \XOR_{T=0}^i {\cal M}_{i,j} \wedge f(T),
\]
where the matrix $\cal M$ is defined in Eq.~\eqref{defM}.

For completeness, we report the expressions for the $\chi^{(8)}_k$ and
$\chi^{(9)}_k$,
\begin{equation}
\label{chi}
\begin{split}
 \chi^{(8)}_1 &= \xi_1 \oplus \xi_3 \oplus \xi_5 \oplus \xi_7,\\
 \chi^{(8)}_2 &= \xi_2 \oplus \xi_3 \oplus \xi_6 \oplus \xi_7,\\
 \chi^{(8)}_3 &= \xi_3 \oplus \xi_7,\\
 \chi^{(8)}_4 &= \xi_4 \oplus \xi_5 \oplus \xi_6 \oplus \xi_7,\\
 \chi^{(8)}_5 &= \xi_5 \oplus \xi_7,\\
 \chi^{(8)}_6 &= \xi_6 \oplus \xi_7,\\
 \chi^{(8)}_7 &= \xi_7,\\
 \chi^{(8)}_8 &= \xi_8;\\
 \chi^{(9)}_1 &= \chi^{(8)}_1 \oplus \xi_9,\\
 \chi^{(9)}_2 &= \chi^{(8)}_2 \quad\dots\quad \chi^{(9)}_8 = \chi^{(8)}_8,\\
 \chi^{(9)}_9 &= \xi_9.
 \end{split}
\end{equation}
Obviously, the $\chi^{(9)}_k$ are only formally similar to the $\chi^{(8)}_k$
 and they are calculated with nine variables.

We note that the normalization condition on the $\chi"_k$ is
\begin{equation}
\sum_{k=0}^n \chi"_k = 1;  \label{sumchi}
\end{equation}
from which $\chi"_0$ can be obtained.

Putting all the stuff together, we need a maximum of 1024 operations
for a generic
rule with eight arguments, and 2304 operations for a generic rule
with nine arguments (if all the operations are explicitly developed);
50 (resp. 57) operations for a generic totalistic rule with
eight (resp. nine)
arguments using directly the
$\xi_i$ of Eq.~\eqref{totalexpansion} and 73 (resp. 82) using the $h_k$ of
Eq.~\eqref{totalistic}. These numbers should be compared with
the $\sim 3000$ operations of the standard disjunctive form
for a function of eight arguments whose truth table is half filled with ones
and with the $\sim 600$
operations required for a totalistic rule with nine arguments as
described in Ref.~\cite{general}.

We can see that the Taylor expansion of a Boolean function allows
a big saving if the function itself depends symmetrically on
the variables (i.e.,\ it is a totalistic function). Sometimes a function
depends in a totalistic way only on part of the variables
(see e.g.,\ the Life rule in the following section). After rearranging
the indices,
\[
	f(x_1,\dots,x_i,x_{i+1},\dots,x_n) =
	  f'(g(x_1,\dots,x_i),x_{i+1},\dots,x_n),
\]
where $g(x_1,\dots,x_i)$ is a totalistic function. If this happens, we have
\[
	f_{(\dots,x_j,\dots,x_k,\dots)} = f_{(\dots,x_k,\dots,x_j,\dots)}\quad
 (j,k \le i)
 \qquad   \forall x \in B_n.
\]

From a computational point of view, $f$ depends symmetrically on
 $x_j$ and $x_k$ if
\begin{equation}
	\OR_{i \in B_n} \left(f_i \oplus f_{i \oplus \delta} \right) = 0,
\label{couple}
\end{equation}
where $\delta=\delta^{(j)} \oplus \delta^{(k)}$. Symmetries among more than
two variables can be obtained via the transition property.

\section{Some applications}
In this section we shall apply the algorithm to some problems,
chosen among the ones that appeared in literature.
Some of them were investigated with efficiency in mind, so they are
supposed to be carefully studied with the aim of reducing the number of
 required operations.

The first example is the totalistic two-dimensional CA
M46789~\cite{notation}. The future value $c'$ of a cell $c$ is determined
by the value most prevalent in its Moore neighborhood (nearest and next to
nearest neighbors, nine variables),
with a twist in case of a marginal majority
or minority. In terms of Eq.~\eqref{totalistic} the rule is defined as
\[
	r_k = \begin{cases}
			1 & \text{if $k = 4,6,7,8,9$;}\\
			0 & \text{otherwise.}
		\end{cases}
\]
The twist in the majority provides a kind of frustration that simulates a
mobile interface according to the Allen-Cahn equation~\cite{interface}.

From the general expression~\eqref{totalexpansion},
we get the simplified expression
\[
   c' = \xi_4\left[1 \oplus \xi_1\left(1 \oplus \xi_2\right)\right] \oplus \xi_8,
\]
for a total of 39 operations.

The second model is the game of Life~\cite{Conway}. This well known game has recently
shown to be a good tool model for the problem of the self-organizing
criticality~\cite{life,Per}.
The evolution rule for Life
depends separately on the sum of the eight nearest and
next-to-nearest neighbors (outer Moore neighborhood),
and on the central cell itself.

The evolution rule can be expressed saying that a dead (zero) cell can become
alive (one) only if it is surrounded by three alive cells, while survival
only occurs if the alive cell is surrounded by two or three alive cells.
Developing first the rule for the central cell $c$, we get
\[
	c' = \chi^{(8)}_3 \oplus \chi^{(8)}_2 \wedge c, 
\]
where $c'$ represent the updated value of the central cell, and the
$\chi^{(8)}_k$ are calculated on the outer Moore neighborhood.

The substitution of  the expressions for the $\chi^{(8)}_k$~\eqref{chi}
and simplification gives
\[
	c' = \xi_2 \wedge \left[c \oplus (1\oplus c)\wedge\xi_1\right] \wedge
\left(1 \oplus \xi_4\right).
\]
As $a \oplus \lnot a \wedge b = a \vee b$ we have
\[
	c' = \xi_2 \wedge \left(1 \oplus \xi_4\right) \wedge \left(c \vee \xi_1\right),
\]
that implies only 33 operations, to be compared with the $\sim 170$
ones reported in Ref.~\cite{general}.

We can also apply the method to non totalistic rules. Let us examine the
Kohring rule~\cite{Khoring} for an FHP~\cite{Frish} gas with obstacles.
The collision rules are the same of the original FHP model, with a set of
four body collisions.
Let us label the six directions in a counterclockwise way
with the indices ranging from one to six. The
operations on the indices are supposed to be modulo six.
All the Boolean quantities are actually elements of some array of  integer
words: we do not consider here the spatial indices.
If the Boolean variable $x_i$ is one, there is an incoming particle on the
site from direction $i$.
Collisions occur for
\[
(x_{j+1}, \dots, x_{j+6}) =\begin{cases}
	(1,0,0,1,0,0), &\text{two particles collisions;}\\
	(1,0,1,0,1,0), &\text{three particles collisions;}\\
	(1,1,0,1,1,0),&\text{four particles collisions.}
\end{cases}
\]
The index $j= 0,\dots,5$ provides for all the rotational invariant configurations.

 After the application of the collision rule,
the variable $y_i$ equal to one means that there is a particle outgoing from
the site with direction $i+3$.	If the particles travel unperturbed, the
updating rule is just $y_i = x_i$.
On each lattice
site there is an additional bit, indicated by $a$,
to code the local conservation of angular momentum. If $a$ is one
and a collision takes place, then all the particles on the site turn
counterclockwise of $\pi/3$ (i.e.,\ $y_i = x_{i-1}$), otherwise the rotation
is clockwise ($y_i = x_{i+1}$). The $a$ bit is reversed each time a collision
takes place.
Finally, a bit $b=0$ (1) indicates the presence (absence) of an obstacle.
The meaning of $b$ is reversed from that in Ref.~\cite{Khoring} for
convenience. In the case $b=0$ no
collision occurs, but the velocities of the particles are reversed,
i.e.,\ $y_i = x_{i+3}$.
For further details about the
implementation we refer to Khoring's work~\cite{Khoring}.

First we want to obtain the expression for a bit $c$ that indicates
the occurrence of a collision.
The two cases of zero and six particles
can eventually be included among the collisions,
without having any consequence.
There are no symmetries among the variables in the truth table of the
collisions (see Eq.~\eqref{couple}),
so it is preferable to divide them into two groups, two and four
particles collisions
in one group, and three particle collisions in the other one.
The first group is characterized by symmetries between $x_i$ and $x_{i+3}$.
Introducing the auxiliary variables $w_i = x_i \oplus x_{i+3}$ (only three
of them are really needed), we get
\[
	c(0,2,4,6) = \chi^{(3)}_0(w1,w2,w3);
\]
where $c(i,j,\dots)$ indicates the contribution to the collision bit by
the $i,j,\dots$ particles collisions, and the $\chi^{(3)}_k$
are the totalistic
characteristic functions for three arguments.
From Eq.~\eqref{sumchi},~\eqref{or} and~\eqref{chi} we obtain
\[
	\lnot c(0,2,4,6) = w1 \vee w2 \vee w3.
\]

Three particles collisions occur when $(x_1,x_3,x_5)$ or $(x_2,x_4,x_6)$
are all zero or one, that is
\[
\begin{split}
  \lnot c&(0,3,6) =\\
	  &= \left[\chi^{(3)}_1(x_1,x_3,x_5) +
	 \chi^{(3)}_2(x_1,x_3,x_5)\right] \vee
	 \left[\chi^{(3)}_1(x_2,x_4,x_6) + \chi^{(3)}_2(x_2,x_4,x_6)\right]\\
   &= \left[\xi_1(x_1,x_3,x_5) \oplus \xi_2(x_1,x_3,x_5)\right] \vee
	 \left[\xi_1(x_2,x_4,x_6) \oplus \xi_2(x_2,x_4,x_6)\right].
\end{split}
\]

Using the property~\eqref{or2} we get
\[
\begin{split}
	\lnot c(0,3,6) &= (x_1 \oplus x_3) \vee (x_3 \oplus x_6) \vee (x_2 \oplus x_4)
\vee (x_4 \oplus x_6);\\
	c &= \lnot \left[ \lnot c(0,2,4,6) \wedge \lnot c(0,3,6)\right].
\end{split}
\]
This expression for the collision bit is equal to that
found in Ref.~\cite{Khoring}.

The expression for the $y_i$ can be though as a function of $a,b,c$.
Developing the expression we get
\[
	y_i= x_{i+3} \oplus b \wedge(x_{i+3} \oplus x_i \oplus z_i),
\]
where
\[
	z_i = \left[x_i \oplus x_{i+1} \oplus (x_{i+1} \oplus x_{i-1}) \wedge a\right]
\wedge c.
\]

Finally, we notice that
\[
	z_i \oplus z_{i+3} = \left[w_i \oplus w_{i+1} \oplus (w_{i+1} \oplus w_{i-1})
\wedge a\right] \wedge c;
\]
but when $c=1$ all the $w_i$ are zero, so $z_i = z_{i+3}$. We
need one more operation to reverse the angular momentum bit in case
of a collision, $a' = c \oplus a$. Taking into
account the common patterns in the expression of $c$,$z_i$ and $y_i$, we
 only need
 35 operations to update all the six velocities, and 14 arrays (six for the
old values of the particles, six for the new values, one for the angular
momentum and one for the collision bits). For comparison, in
Ref.~\cite{Khoring} the algorithm needs 74 operations and 16 arrays.

Incidentally, we observe that only six arrays are really needed to store
the configuration, without the needing of a translational phase. Indeed,
the rule only implies an (eventual) exchange of the particles among the planes
(the RAP1 machine~\cite{RAP1} is based on this consideration). The translation
of the particles can be taken into account by a logical shift of the origin
of the arrays. The procedure is still vectorizable, but the mapping between
the lattice and the arrays of integer words is indeed more complex, so maybe
it is not worth doing the efforts unless perhaps for a dedicated hardware.

The last example involves a probabilistic totalistic CA for
the simulation of the Ising model~\cite{Herrmann}.
The rule depends in a totalistic way on the outer Von Neumann neighborhood (the
north, east, south and west neighbors). The rule can be
expressed as
\[
	c' = \sum_{k=0}^4 r_k \wedge \chi^{(4)}_k,
\]
where the $r_k$ are random bits equal to one with predefined
 probabilities $p_k =p(r_k)$.
Building the $\chi^{(4)}_k$ from the $\xi_i$ as in Eq.~\eqref{chi},
we get the quoted
result of 22 Boolean operations and four arithmetic summations. Writing down
the RSE~\eqref{rse}, we have
\[
	c' = r_0 \oplus \XOR_{i=1}^4 s_i \wedge \xi_i,
\]
where the $s_i$ are random bits with probability $p(s_i)$, obtained as
\[
\begin{split}
	s_1 &= r_0 \oplus r_1, \\
	s_2 &= r_0 \oplus r_2, \\
	s_3 &= r_0 \oplus r_1 \oplus r_2 \oplus r_3,\\
	s_4 &= r_0 \oplus r_4;\\
\end{split}
\]
and considering that $p(a \oplus b) = p(a) + p(b) - 2 p(a) p(b)$.

Using the approach described above, we need nine operations to build up the
$\xi_i$ and eight operations for $c'$.

\section{Conclusions}
In this work we extend and complete the notion of the derivatives
of a Boolean function already
introduced in Ref.~\cite{Vichniac}. We are thus able to derive the Taylor and
MacLaurin series of a Boolean function. The latter represent the
ring sum expansion for a Boolean function, which is more compact
than the canonical conjunctive and disjunctive forms. Moreover, for
totalistic functions (i.e., for functions completely symmetric in their
arguments) very compact expressions are found. These ideas have wide
applications in the development of faster algorithms, in particular for
cellular automata simulations, and in the design of digital circuitry.
As examples of physical applications, we analyze
already published optimized algorithms, involving
both deterministic and stochastic
automata. We found that our procedure generally leads
to more compact expressions. We think that the Boolean derivative
is not limited to the minimization of
Boolean functions. Work is in progress
about the connection between Boolean derivatives and the chaotic
properties of cellular automata, possibly leading to the definition of
Lyapunov exponents for discrete systems.

\section*{acknowledgements}
 We are grateful to G. Vichniac, R. Livi and S. Ruffo
for fruitful discussions. We also
acknowledge the Institute for Scientific Interchange Foundation (Torino,
Italy) where this work was first started in the framework of the
workshop {\sl Complexity and Evolution}.

\end{document}